# Dirac Fermions and Superconductivity in Homologous Structures $(Ag_xPb_{1-x}Se)_5(Bi_2Se_3)_{3m, m = 1,2}$

L. Fang[1, *], C. C. Stoumpos[2], Y. Jia[3], A. Glatz[2,4], D. Y. Chung[2], H. Claus[2], U. Welp[2], W. K. Kwok[2] and M. G. Kanatzidis[1,2, *]

[1] Department of Chemistry, Northwestern University, Illinois 60208
[2] Materials Science Division, Argonne National Laboratory, Argonne, Illinois 60565, United States
[3] Department of Physics and Astronomy, Northwestern University, Evanston, Illinois 60208, USA
[4] Department of Physics, Northern Illinois University, DeKalb, Illinois 60115, United States

**A newly discovered topological insulator (TI) $(Ag_xPb_{1-x}Se)_5(Bi_2Se_3)_{3m, m=2}$, has a band gap of 0.5 eV, the largest value ever reported in topological insulators (TIs). We present a magneto-conductivity study of the Dirac electrons of this compound in the quantum diffusion regime. Two-dimensional weak anti-localization was observed and identified as destructive interference caused by the Berry's phase of this topological state. We find that the phase coherence length of the Dirac electrons is independent of doping and disorder levels. This provides proof for the absence of backscattering arising from the protection of time reversal invariance in TI $(Ag_xPb_{1-x}Se)_5(Bi_2Se_3)_{3m, m=2}$. We further report that the homologous compound $(Ag_xPb_{1-x}Se)_5(Bi_2Se_3)_{3m, m=1}$ is a superconductor with a transition temperature $T_c$ = 1.7 K. The related structures of these two phases allow lateral intergrowth of crystals to occur naturally, offering an opportunity to observe the Majorana Fermion state at the boundary of two inter-grown crystals.**

* Corresponding authors
m-kanatzidis@northwestern.edu
lei.fang@northwestern.edu

The observation of quantum phenomena in topological insulators (TIs) is closely related to the discovery of new materials [1-4]. Numerous advances on TIs [5-9], both theoretical and experimental, were based on two binary bismuth compounds, $Bi_2Se_3$ and $Bi_2Te_3$. Material tailoring of $Bi_2Se_3$ with other structures can produce new phases that host topological states [11, 12]. One example is the homologous series of layered structures of $(PbSe)_5(Bi_2Se_3)_{3m, m=1, 2}$ [13], and the presence of a Dirac cone in the $m$=2 phase coupled with a large band gap of 0.5 eV [14]. The $m$=2 phase is of high interest because this largest ever-known band gap may give rise to unusual properties. Knowledge about this material, however, is very limited. Here we report on the $m$=1 and 2 phases of a novel topological insulator heterostructure, $(Ag_xPb_{1-x}Se)_5(Bi_2Se_3)_{3m}$ featuring a large band gap (0.5 eV), and investigate the magneto-conductivity properties of Dirac electrons in the quantum diffusion regime. A two-dimensional (2D) weak anti-localization (WAL) was observed and identified as destructive interference caused by the Berry's phase of the topological state [6]. We find that the phase coherence length of the Dirac electrons is independent of doping and disorder levels. This provides evidence for the absence of backscattering due to the protection by time reversal invariance in the $m$=2 phase of this TI. Moreover, we show that the $m$=1 phase is a superconductor with a critical temperature of $T_c$ = 1.7 K, and a strong spin-orbit-coupling (SOC) in the normal state. The homologous structures of these two phases allow intergrowth of crystals to occur naturally. We thus propose a possibility to observe the Majorana Fermion state at the boundary of two inter-grown crystals.

The structures of $(PbSe)_5(Bi_2Se_3)_{3m, m=1, 2}$ are stacks of layers of two building blocks [PbSe] and [$Bi_2Se_3$], as shown in Fig. 1(b) and (c). The value of $m$ designates the number of [$Bi_2Se_3$] sheets that are sandwiched by consecutive [PbSe] slabs. The [$Bi_2Se_3$] sheet is hexagonal and is the same as a quintuple layer (QL) of TI $Bi_2Se_3$. The [PbSe] sheet has tetragonal symmetry and is a two-atom thick slice from the face-centered-cubic (FCC) structured PbSe. The lattice mismatch between [$Bi_2Se_3$] and [PbSe] produces a stress that strongly distorts the slab of [PbSe] along the $c$-axis for both homologous phases. The $m$=1 phase features a large number of chemical bonds with a length of ~3.25 Å between the [PbSe] and [$Bi_2Se_3$] sheets creating a 3D structure. The $m$=2 phase, the two adjacent QLs II and III are connected by van der Waals bonding (see Fig. 1(c)). This weak bonding relaxes the stress and preserves the uniformity of the [$Bi_2Se_3$] sheets. Crystal cleaving always occurs along the van der Waals plane and exposes an ideal surface of $Bi_2Se_3$ with Dirac Fermions residing on it. By contrast, crystal cleavage in the $m$=1 phase, breaks the chemical bonding between [PbSe] and [$Bi_2Se_3$], likely leaving a large number of dangling bonds on top of the [$Bi_2Se_3$] slab. In this regard, the cleaved surface of the *m* = 1 phase may suppress any topological states because of the dangling bonds creating unpaired electrons and the stress from the structural distortion in the [$Bi_2Se_3$] QL. As a matter of fact, angle resolved photoemission spectroscopy (ARPES) measurements observed a Dirac cone in $(PbSe)_5(Bi_2Se_3)_6$, in contrast to a regular band structure for $(PbSe)_5(Bi_2Se_3)_3$ [14]. To explain the exceedingly different electronic structures, a scenario of topological phase transition has been proposed [9, 14]. Here we show that transition metals such as silver can replace in part Pb atoms in the structure up to levels of 25% while maintaining the TI charcter in *m*=2 but inducing superconductivity in *m*=1 phase.

Our single crystal *X*-ray diffraction analysis detected unusually weak thermal displacement factors at the Se sites, revealing Se vacancies in the *m*=2 phase (see supplementary information). These vacancies donate electrons and elevate the Fermi level ($E_F$) toward the bottom of the conduction band. Both the temperature dependent resistivity of $(PbSe)_5(Bi_2Se_3)_6$ and Ag-doped $(Ag_{0.2}Pb_{0.8}Se)_5(Bi_2Se_3)_6$ exhibit semimetal behavior and moderately high residual resistivity, as shown in Fig. 1(d). The Hall resistance of the *m*=2 phase exhibits a non-linear field dependence (Fig. 2(a) inset), suggesting that multiple types of carriers contribute to the transport process, which, in our case, are the Dirac fermions on the surface and the conventional electrons in the bulk. This assertion is supported by magneto-conductivity measurements discussed below.

At low temperatures and in low fields, a TI-related weak anti-localization (WAL) effect [16-21], is observed in $(Ag_{1-x}Pb_xSe)_5(Bi_2Se_3)_6$. In the quantum diffusion regime, the phase coherence length $l_\varphi$ of the electrons is much longer than the mean free path $l_e$. The constructive phase interference localizes the electrons and leads to a correction to the conductivity. This quantum corrected conductivity is referred to as weak localization [22]. In TI, the π-Berry's phase rooted in the Dirac cone destroys the phase interference and gives rise to a negative magneto-conductivity, resulting in WAL [23]. The WAL induced magneto-conductance near zero field can be experimentally obtained from *ΔG(*B*)= 1/R(*B*)-1/R(0)*, and its angular dependence can be obtained from *ΔG*(B, *θ*)=1/*R*(B, *θ*)-1/*R*(B, 90º) [18], where θ is the angle between the field and the *c*-axis; *R*(*B*, 90º) is the resistance when the field is parallel to the in-plane current.

Figure 2 (a) depicts the field dependent magneto-conductance of a bulk crystal $(PbSe)_5(Bi_2Se_3)_6$ at 2 K in various magnetic field directions. The sharp cusps of the negative magneto-conductance near zero field manifest the WAL effect in $(PbSe)_5(Bi_2Se_3)_6$. The magneto-conductance measured at various angles collapses onto one single curve when using the normal component of the applied field as the x-coordinate. This scaling unveils the 2D character of the WAL. 2D WAL was also observed in *m*=2 phase with a doping level x=0.2 (see supplementary information). By analogy to $Bi_2Se_3$, we may attribute the WAL to the 2D topological state on surface. However, due to the strong SOC and highly anisotropic structure/geometry, 2D WAL could possibly occur in the bulk electrons as well [22]. As a matter of fact, the origin of the WAL in $Bi_2Se_3$ is still under debate. We suggest that this ambiguousness can be addressed by the measurement of WAL in a reference sample. This reference sample has to be topologically trivial, but possess similar structure and SOC strength as that of the TI of interest. In our case, the topologically trivial phase $(PbSe)_5(Bi_2Se_3)_3$ is the ideal reference for the TI $(PbSe)_5(Bi_2Se_3)_6$ because of their resemblance in composition and structure. Fig. 2(b) shows the angle dependent magneto-conductance of a $(Ag_{0.2}Pb_{0.8}Se)_5(Bi_2Se_3)_3$ crystal at 2 K. The normalized conductance has a negative field dependence and a cusp shape, analogous to the WAL observed in TI *m*=2 phase. However, the magneto-conductance of this phase is independent of field direction. This 3D feature is distinct from the 2D characteristic of a topological surface state. The WAL in the *m*=1 phase is thus ascribed to the destructive interference of bulk electrons caused by strong SOC.

Our hypothesis of the similar SOC strength of the two homologous phases is supported by the comparison of their in-plane magneto-conductivity σ(B, θ=90°). As shown in the Fig. 3(c), σ(B, θ=90°) of *m*=2 phase, silver doped *m*=2 phase and silver doped *m*=1 phase are scaled by one conductivity quantum, $e^2/\pi h$ [24]. These curves are virtually identical, strongly suggesting a SOC nature of the in-plane WAL of the TI phase. In order to quantitatively understand the 3D WAL effect, we fitted the magneto-conductivity in Fig. 3(c) by using the theory of the 3D WAL. The correction to the conductivity $\sigma_{3D}$ is written as [25]

$$\Delta\sigma_{3D}(B) = \frac{e^2}{4\pi^2\hbar}\sqrt{\frac{B}{B_\varphi}}\left[3F\left(\left[1+\frac{4\tau_\varphi}{\tau_{SO}}\right]\frac{B_\varphi}{B}\right) - F\left(\frac{B_\varphi}{B}\right)\right] \quad (1)$$

$$F(x) = \sum_{k=0}^{\infty}\left[2(\sqrt{k+1+x} - \sqrt{k+x}) - 1/\sqrt{k+\frac{1}{2}+x}\right]$$

, where $B_\varphi = \hbar/(4el_\varphi^2)$ is characteristic field and $l_\varphi$ is phase coherence length. $\tau_\varphi$ and $\tau_{so}$ are inelastic scattering time and spin-orbit time, respectively. When $\tau_\varphi \gg \tau_{so}$, Eq.ation (1) simplifies to

$$\Delta\sigma_{3D}(B) \cong -\frac{e^2}{4\pi^2\hbar}\sqrt{\frac{B}{B_\varphi}}F\left(\frac{B_\varphi}{B}\right). \tag{2}$$

σ(B, θ=90°) of the studied three samples can be fitted by Eq. (2) with a fitting parameter $B_{\varphi, 3D}$=0.013 T. The obtained $l_\varphi \approx$ 110 nm is far greater than the c-axis length of the TI *m*=2 phase, suggesting a strong inter-layer coupling of the SOC effect in this phase. The 3D fitting, in addition to the identical magneto-conductivity with a non-TI sample, unambiguously verifies that the in-plane WAL of the TI *m*=2 phase arises from the strong SOC. This 3D SOC also contributes the corrected-conductivity for field applied out-of-plane. In this regard, σ(B, θ=0°) of the TI *m*=2 phase is a corrected conductivity composed by both the SOC effect and the destructive interference from Berry's phase.

The phase coherence length of the Berry's phase induced 2D WAL can be obtained by using the Hikami-Larkin-Nagaoka (HLN) theory. Assuming that the inelastic scattering time is much longer than both spin-orbit and elastic scattering time, $\tau_\varphi \gg \tau_{so}$, $\tau_\varphi \gg \tau_e$, the quantum correction to the conductivity can be written as [26],

$$\Delta\sigma(B) \cong -\alpha\frac{e^2}{\pi h}\left[\Psi\left(\frac{1}{2}+\frac{B_\varphi}{B}\right) - \ln\left(\frac{B_\varphi}{B}\right)\right] \tag{3}$$

where α=1/2. $\Psi(z)$ is the digamma function. $l_\varphi = \sqrt{D\tau_\varphi}$, and *D* is the diffusion constant $D = \frac{1}{d}v_F^2\tau_e$, where $v_F$ is the Fermi velocity and *d* represents the sample dimension. ℏ is Planck constant divided by 2π. Fig. 3(d) shows the Δσ(B) of *m*=2 phase and the HLN fitting. The obtained $B_{\varphi, 2D}$ is about 0.007 T and $l_\varphi \approx$160 nm, comparable to the reported value of 310 nm in TI $Bi_2Te_3$. In the same figure, WAL of another crystals of the *m*=2 phase and one doped specimen (*m*=2, x=0.2) are plotted. It is striking that the WAL of different samples and doping levels can be scaled onto one curve over a broad field range from -0.4 to 0.4 T. This scaling unveils that the characteristic field and the phase coherence length $l_\varphi$ are independent of the specifics of the sample. To obtain a constant value of $l_\varphi$, both *D* and $\tau_\varphi$ have to be invariant. This observation can only be understood in the context of topological insulators. Since the dispersion relation is linear, $E(k)\sim k$, $v_F$ of the Dirac electrons is independent of $E_F$. $\tau_e$ is retained because of backscattering is forbidden due to time reversal invariance (TRI)[27-28]. $\tau_\varphi$ is not altered unless the Berry's phase is destroyed by magnetic impurities. Our experiments thus provide evidence for TRI in $(PbSe)_5(Bi_2Se_3)_6$. In addition to this immunity to backscattering, our analysis also sheds light on the corrected conductance. The value $B_{\varphi, 2D}$ (0.007 T) is rather close to that of $B_{\varphi, 3D}$ (0.013 T), revealing that the conducting corrections due to the Berry's phase and the SOC effect in TI $(PbSe)_5(Bi_2Se_3)_6$ are comparable. This observation, if it can be generalized to other TI systems, could offer new insights on the origin of the WAL in TIs.

We were not able to induce superconductivity in $(PbSe)_5(Bi_2Se_3)_6$ by doping with silver and other transition metals (see supplementary information). However, we found that the Ag-doped $m=1$ phase $(Ag_xPb_{1-x}Se)_5(Bi_2Se_3)_3$ is superconducting. As shown in Fig. 3(a), the resistance of $(Ag_{0.2}Pb_{0.8}Se)_5(Bi_2Se_3)_3$ starts to drop at 1.6 K and zero resistance is reached at 0.55 K. Bulk superconductivity was observed for different doping levels. Fig. 3 (b) presents the magnetization measurements of $(Ag_xPb_{1-x}Se)_5(Bi_2Se_3)_3$ (x=0.1~0.25). At doping levels x < 0.1, the diamagnetic signal is extremely weak. At doping of x > 0.25, the structure of $(Ag_xPb_{1-x}Se)_5(Bi_2Se_3)_3$ becomes unstable. The highest $T_C$ is 1.7K. In order to determine the upper critical fields ($H_{C2}$) and superconducting anisotropy (Γ), we suppressed the superconductivity by applying magnetic fields in plane and out of plane, as shown in Fig. 3(c) and (d). Using 90% of the normal-state resistance as a criterion, we mapped out the superconducting phase diagram of $(Ag_xPb_{1-x}Se)_5(Bi_2Se_3)_3$, shown in the inset of Fig. 3(a). At T≤ $T_C$, $\frac{dH_{c2}^c}{dT} \approx -0.67\ T/K$ and $\frac{dH_{c2}^{ab}}{dT} \approx -1\ T/K$. Using the Werthamer-Helfand-Hohenberg formula [29] $H_{C2}(0)=-0.693\ T_C\ (dH_{C2}/dT)_{T=T_c}$, the zero temperature upper critical fields are $H_{c2}^c \approx 0.74\ T$ and $H_{c2}^{ab} \approx 1.1\ T$. The Ginzburg-Landau in-plane coherence length can be obtained from $\xi_0^{GL} = \sqrt{\phi_0/2\pi H_{c2}^c}$ =21.7 nm, where $\Phi_0$ is the flux quantum. The superconducting anisotropy [30] is written as $\Gamma = \sqrt{\frac{m_c}{m_{ab}}} = \frac{\xi_{ab}}{\xi_c} = \frac{H_{c2}^{ab}}{H_{c2}^c}$ = 1.5, where $m_c/m_{ab}$ is the ratio of effective mass matrix elements. The near-isotropic characteristics of $H_{C2}$ is consistent with our structure determination and the 3D feature of the SOC-induced WAL.

The superconducting $(Ag_{0.2}Pb_{0.8}Se)_5(Bi_2Se_3)_3$ may provide an opportunity to detect Majorana Fermions at the interface of $m=1$ and $m=2$ phases. Because of the very close matching of in-plane lattice parameters, the $(Ag_{0.2}Pb_{0.8}Se)_5(Bi_2Se_3)_3$ crystals can be easily inter-grown with crystals of $(Ag_{0.2}Pb_{0.8}Se)_5(Bi_2Se_3)_6$ with limited strain. This creates a natural interface between a superconductor and a topological insulator. Taking advantage of the proximity effect, superconductivity can transform some of the Dirac Fermions of the topological state into an Andreev surface state [31]. Hence, Majorana Fermions could be observable at the boundary of the $(Ag_{0.2}Pb_{0.8}Se)_5(Bi_2Se_3)_3$ and the $(Ag_{0.2}Pb_{0.8}Se)_5(Bi_2Se_3)_6$ domains. The interface of intergown samples was imaged with polarized light on a cleaved surface of a superconducting crystal. As shown in the Fig. 4(a), two domains with distinct cleavage morphologies were observed to grow together. The domain boundary is surprisingly uniform and may allow local detection techniques.

In conclusion, we report the crystal structures of the Ag-doped TI $(PbSe)_5(Bi_2Se_3)_{3m, m=2}$ and its homologue $(PbSe)_5(Bi_2Se_3)_{3m, m=1}$. Angular dependent measurements of WAL allow us to separate effects arising from SOC from those arising from 2D surface states. For the superconducting m=1 phase, the superconducting gap Δ of $(Ag_xPb_{1-x}Se)_5(Bi_2Se_3)_3$(x=0.2) can be estimated by equation $\Delta = \frac{\hbar v_F}{\pi \xi_0^{BCS}}$, where $v_F$ is Fermi velocity and is assigned to be $10^5$ m/s, a value for the regular solid state materials. In our case, Bardeen-Cooper-Schrieffer (*BCS*) coherence $\xi_0^{BCS}$ is close to the Ginzburg-Landau in-plane coherence length. The obtained gap value of ~ 1 meV, indicates a weak pairing potential in this superconductor. The strong SOC in *m*=1 phase could induce unconventional superconductivity, for instance, nodes in the superconducting gap [32]. The recently discovered topological superconductor $Cu_xBi_2Se_3$ also exhibits signatures of odd parity of pairing symmetry [33, 34]. Although the energy of the band splitting of spin-up and spin-down has not been determined in $(Ag_xPb_{1-x}Se)_5(Bi_2Se_3)_3$(x=0.2), a striking band splitting with energy about 180 meV was found in TI $Bi_2Se_3$ by spin-polarized ARPES [35]. This energy scale exceeds the gap values of most type II superconductors. Assuming a comparable band splitting in the superconducting *m*=1 phase, unconventional paring may happen. Experimental determination of this superconducting pairing symmetry is beyond the scope of this paper. Future investigations, such as tunneling spectroscopy, will be performed to clarify this issue.

Experiments

**Crystal growth of $(Ag_xPb_{1-x}Se)_5(Bi_2Se_3)_6$** Precursor $PbBi_4Se_7$ was homemade by solid-state synthesis using high-purity elements (American Elements, purity>99.999%). Silver (purity>99.999%) and precursor were weighed in a nominal composition $Ag_xPbBi_4Se_7$(x=0~0.3). The mixture was sealed in an evacuated quartz tube and loaded in a tube furnace. The furnace was first heated to 950 °C in 12 h and kept at this temperature for 12 h. Subsequently, the furnace was slowly cooled down to 650 °C in 80 h and shut off. Shiny, centimeter-sized crystals were produced. Intergrowth of two homologous phases was generally observed for crystals with size larger than half millimeter. However, phase-pure single crystals with size about 300~400 μm can be separated from the ingot.

**Crystal growth of $(Ag_xPb_{1-x}Se)_5(Bi_2Se_3)_3$**  Ag, PbSe (homemade) and $Bi_2Se_3$ (homemade) were weighted in a nominal composition $Ag_xPb_5Bi_6Se_{14}$ and ground thoroughly. The temperature profile for crystal growth is the same as that of $(Ag_xPb_{1-x}Se)_5(Bi_2Se_3)_6$. Sizable single crystals can be obtained in a narrow doping range from x=0.2~0.3.

**Structure determination and characterization**  Single crystal diffraction was carried out on a STOE IPDS II Diffractometer. Structure solution and refinement was conducted using the SHELX-97. Silver concentration was determined by EDX on a Hitachi S-3400N VP-SEM.

Crystal data:

1) $(PbSe)_5(Bi_2Se_3)_6$, Monoclinic, Space group C2/m, c = 52.918(10) Å, γ = 90.00°, b = 4.1774(8) Å, β = 107.224(3)°, a = 21.551(4) Å, α = 90.00°, T=293 K, V=4550.4(15) Å$^3$, Z=4, Reflections collected 26308, Independent reflections 6220 [$R_{int}$=0.0971], Completeness to θ = 29.00°, 90.8%, refinement method, Full-matrix least-squares on F$^2$, Goodness-of-fit, 1.054, Final R indices [>2σ(I)] $R_{obs}$ = 0.0708, R indices [all data] $R_{all}$ = 0.1321.

2) $(Ag_xPb_{1-x}Se)_5(Bi_2Se_3)_6$, Monoclinic, Space group C2/m, c = 52.927(4) Å, γ = 90.00°, b = 4.1747(2) Å, β = 107.423(6)°, a = 21.5211(17) Å, α = 90.00°, T=293 K, V=4537.0(5) Å$^3$, Z=4, Reflections collected 22210, Independent reflections 6894 [$R_{int}$=0.1299], Completeness to θ = 29.00°, 99.3%, refinement method, Full-matrix least-squares on F$^2$, Goodness-of-fit, 0.915, Final R indices [>2σ(I)] $R_{obs}$ = 0.0589, R indices [all data] $R_{all}$ = 0.1504.

3) $(PbSe)_5(Bi_2Se_3)_3$, Monoclinic, Space group P2$_1$/m, c = 15.9840(12) Å, γ = 90.00°, b = 4.1915(3) Å, β = 97.4750(10)°, a = 21.4727(16) Å, α = 90.00°, T=293 K, V=1426.38(18) Å$^3$, Z=2, Reflections collected 7977, Independent reflections 3359 [$R_{int}$=0.1129], Completeness to θ = 29.00°, 93.1%, refinement method, Full-matrix least-squares on F$^2$, Goodness-of-fit, 0.859, Final R indices [>2σ(I)] $R_{obs}$ = 0.0624, R indices [all data] $R_{all}$ = 0.1091. [36]

**Magnetization measurements**  Magnetization measurements were carried out using a homemade low-field (0.1 G) SQUID magnetometer operating at temperatures down to 1.2K. The measurements were performed on warming after initially cooling in zero field (< 5mG) to 1.2 K.

**Low temperature transport measurements**  Single crystals with freshly cleaved, mirror-like surfaces were cut into rectangular shapes. Ohmic contacts were fashioned into a Hall-bar geometry. Sliver paste with nano-sized Ag particles was painted on both the surfaces and the edges. WAL was measured in a LHe$^4$ variable temperature cryostat equipped with a triple-axis vector magnet system (AMI) for accurate field orientations. Superconducting transitions were measured in a LHe$^3$ cryostat.

**Acknowledgements**:

We are grateful to L. Bouchard and Kang L. Wang for useful discussions. This research was supported by the Defense Advanced Research Project Agency (DARPA), Award No. N66001-12-1-4034. Transport and magnetizations measurements were supported by the Department of Energy, Office of Basic Energy Sciences, under Contract No. DE-AC02-06CH11357 (DY, CCS, YJ, HC, AG, UW, WKK).

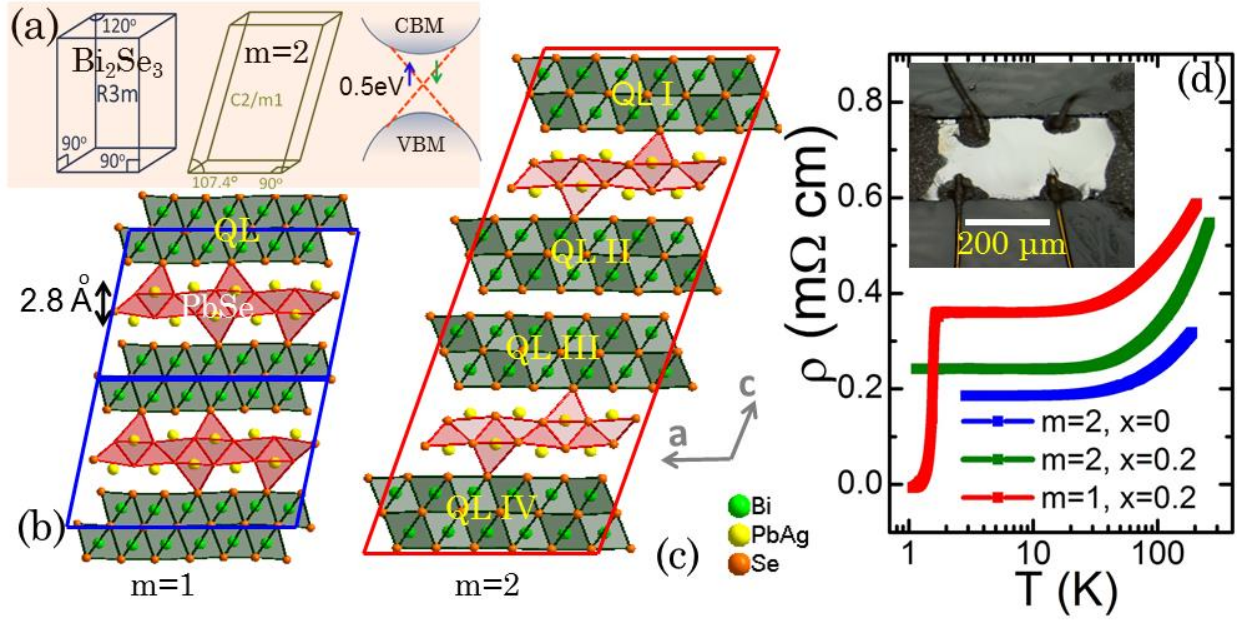

Figure 1

(a) Schematic pictures of the crystal symmetry of topological insulators $Bi_2Se_3$ and $(PbSe)_5(Bi_2Se_3)_6$ and of a Dirac cone in the band structure of the $m$=2 phase. (b) Structure of two unit cells of the $m$=1 phase. The [PbSe] slab is distorted due to the lattices mismatch between [PbSe] and [$Bi_2Se_3$]. The thickness of [PbSe] is ~2.8 Å. [PbSe] and [$Bi_2Se_3$] are connected by chemical bonds. (c) Structure of one unit cell of the $m$=2 phase. Bonds between QL-II and QL-III are van der Waals type. Silver can replace Pb ions in the [PbSe] slabs. (d) Temperature dependent resistivity of $(Ag_xPb_{1-x}Se)_5(Bi_2Se_3)_6$ (x=0, 0.2) and $(Ag_{0.2}Pb_{0.8}Se)_5(Bi_2Se_3)_3$. The inset shows a cleaved single crystal of the $m$=2 phase with Ohmic contacts in a Hall bar geometry.

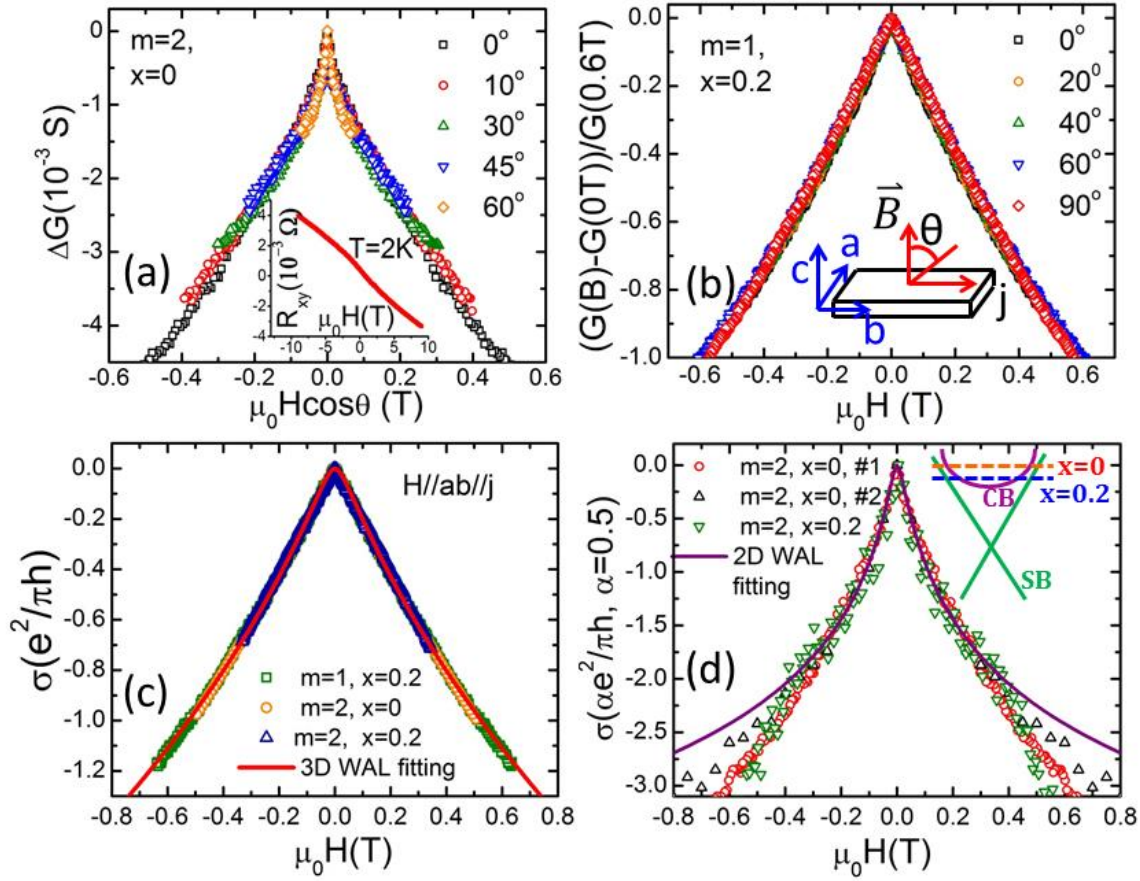

Figure 2

(a) Quantum correction on the conductance ΔG of $(PbSe)_5(Bi_2Se_3)_6$ at 2 K. The sharp cusp at low fields is a fingerprint of the weak anti-localization effect. ΔG at different angles can be scaled–by $H\cos\theta$, revealing the 2D characteristic of the WAL. The inset is the field dependent Hall resistance of $(PbSe)_5(Bi_2Se_3)_6$. The non-linear Hall resistance is caused by the coexistence of two conducting channels, topological surface channel and bulk conductance. (b) Field dependence of conductance of the topological-trivial $m=1$ phase shows a WAL effect as well. The normalized $G$ with applied fields at different tilted angles superimpose onto each other, indicating a 3D behavior of this WAL. WAL of the $m=1$ phase is ascribed to the strong SOC effect in a bulk crystal. The inset is a schematic picture of angle dependent magneto-conductivity measurement. (c) Magneto-conductivity of $m=2$ phase, doped $m=2$ phase and doped $m=1$ phase are scaled by one conductivity quantum, $e^2/\pi h$. The solid curve is a fit using the 3D weak (anti)localization theory. (d) $\Delta G/(e^2/\pi h)$ of two samples (sample#1, #2) of the TI $(PbSe)_5(Bi_2Se_3)_6$ and of a silver doping ($m=2$, x=0.2) crystal, in magnetic fields along the c-axis. The solid curve is a simulation using the HLN equation with a pre-factor $\alpha=1/2$. Inset schematically shows the coexistence of bulk conductance band (CB) and surface band (SB). Silver substitution donates holes and decreases the Fermi energy ($E_F$) to lower energy.

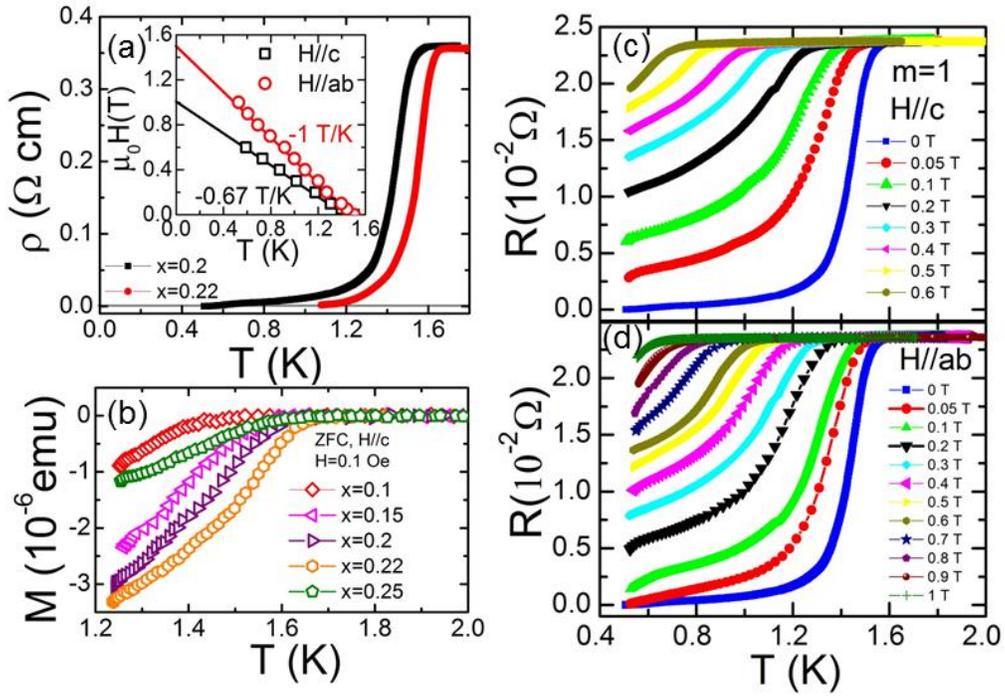

Figure 3

(a) Temperature dependent resistivity of $(Ag_xPb_{1-x}Se)_5(Bi_2Se_3)_3$ (x=0.2, 0.22). Both sample show superconducting transitions below 1.6 K and zero resistances at lower temperatures. The inset is a $H_{c2}$–plot for doping x=0.2 with fields applied in-plane and out-of-plane. The straight solid lines are guides to the eye. (b) Temperature dependent magnetization of crystals $(Ag_xPb_{1-x}Se)_5(Bi_2Se_3)_3$ (x=0.1~0.25). $T_c$ varies with the sliver doping. (c) and (d) are temperature dependent resistance of the superconducting sample of $(Ag_{0.2}Pb_{0.8}Se)_5(Bi_2Se_3)_3$. Both out-of-plane and in-plane magnetic fields were applied to suppress the superconductivity.

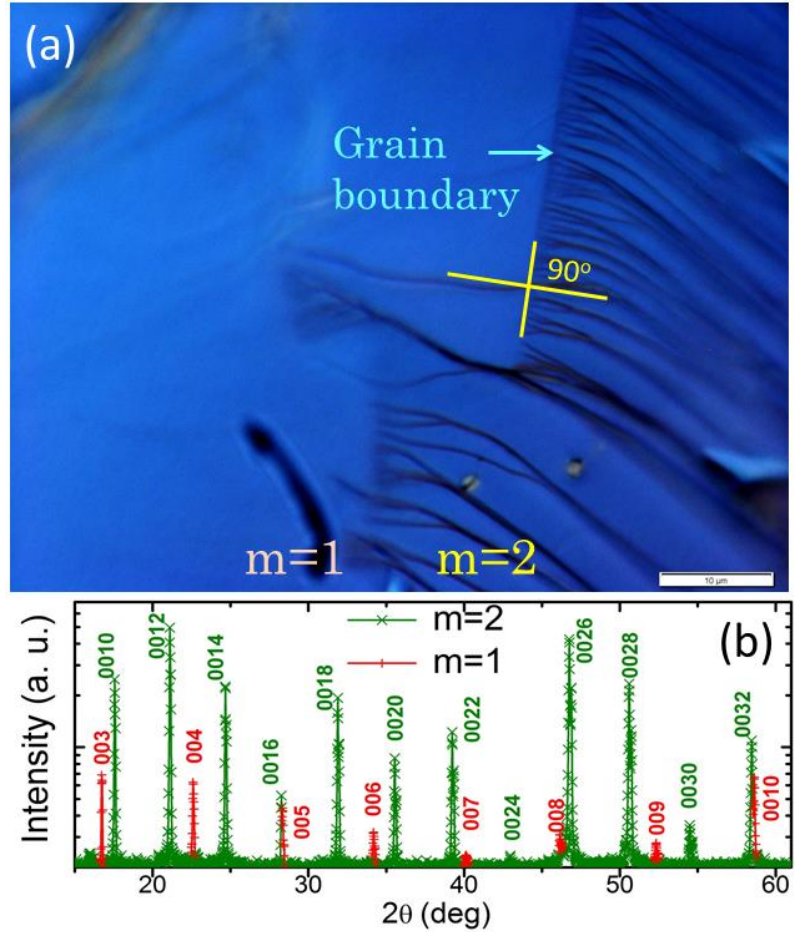

Figure 4

(a) Optical image of a cleaved superconducting crystal. Polarized light was applied to enhance the contrast between different crystalline domains. Grain boundaries divide the crystal in two parts. The right part exhibits a more cleavable characteristic than the part on the left. This distinct topography indicates the different structures of these two domains. The domain angles, as shown as the yellow cross, are ~$90^o$ , consistent with the solved crystallographic data of $(PbSe)_5(Bi_2Se_3)_{3m, m=1, 2}$. The right domain is ascribed to the *m*=2 phase and the left part is *m*=1 phase. (b) X-ray diffraction of a superconducting crystal along the *c*-axis. The two sets of (*00l*) peaks unveil that two phases inter-grow in the *ab* plane, consisting well with the image in (a).